\documentclass[oribibl]{llncs}

\newtheorem{observation}{Observation}

\begin{document}

\title{Position Auctions with Externalities and Brand Effects\thanks{September 16, 2014.  A shorter version of this paper will appear in WINE 2014.}}
\author{Patrick Hummel\inst{1} \and R. Preston McAfee\inst{2}}
\institute{Google Inc. \email{phummel@google.com} \and Microsoft Corp. \email{preston@mcafee.cc}}

\maketitle

\begin{abstract}
This paper presents models for predicted click-through rates in position auctions that take into account two possibilities that are not normally considered---that the identities of ads shown in other positions may affect the probability that an ad in a particular position receives a click (externalities) and that some ads may be less adversely affected by being shown in a lower position than others (brand effects).  We present a general axiomatic methodology for how click probabilities are affected by the qualities of the ads in the other positions, and illustrate that using these axioms will increase revenue as long as higher quality ads tend to be ranked ahead of lower quality ads.  We also present appropriate algorithms for selecting the optimal allocation of ads when predicted click-through rates are governed by either the models of externalities or brand effects that we consider.  Finally, we analyze the performance of a greedy algorithm of ranking the ads by their expected cost-per-1000-impressions bids when the true click-through rates are governed by our model of predicted click-through rates with brand effects and illustrate that such an algorithm will potentially cost as much as half of the total possible social welfare.
\end{abstract}

\section{Introduction}

Sponsored search advertising is big business, with major search engines collectively generating tens of billions of dollars of revenue per year.  In sponsored search auctions, advertisements appear alongside search results from major search engines in a variety of positions on the page, some of which are more prominent and thus more likely to be clicked than others.  In both academic work and practice, it is standard to model each position as having some quality score that reflects the relative probability that an ad will receive a click in that position and then ranking the ads by a product of their bid, the maximum amount the advertisers will pay per click, and a quality score, which reflects the probability an ad will receive a click if the advertiser is shown in the top position. 

Although it is almost universal to assume that an ad's click probability is a product of the quality score of the ad and a quality score of a position, there are several reasons that this formulation may be suboptimal.  First, this formulation implicitly assumes that the probability an ad receives a click in a given position is independent of the identities of the other ads that are shown on the page.  However, this assumption is unlikely to hold in practice.  \cite{Craswell08} notes empirically that many consumers are likely to search for what they are looking for by beginning their search at the top of the page and ceasing to search after they have found what they are looking for, and \cite{Jeziorski14} presents empirical evidence that ads impose large negative externalities on other ads by virtue of the fact that the ads can be substitutes for one another.  Consequently, placing higher quality ads at the top of the page decreases the probability that a user clicks on other ads.

Another potential problem with the standard formulation of click probabilities in sponsored search auctions is that this formulation also implies that the percentage decrease in the probability that any given ad will receive a click as a result of being placed in a lower position is the same for all ads.  But this relationship is also unlikely to hold empirically.  For instance, \cite{Jerath11} notes that if there are some consumers who simply gravitate towards the highest-quality company regardless of where that company's ads are on the page, whereas other consumers simply gravitate towards the top of the page first, then ads from well-known companies will be less adversely affected by being shown in a lower position on the page. \cite{Jerath11} also provides empirical support for this.

How then, might one incorporate these possibilities into sponsored search auctions to choose a more efficient allocation of ads?  This paper presents methods for achieving this goal.  We begin by presenting a general axiomatic model of predicted click-through rates when the probability an ad receives a click may depend not only on the quality score of the ad and the position in question, but also on the quality scores of the other ads that are shown in the other positions.  We analyze the properties of this axiomatic formulation, and illustrate that as long as higher quality ads are typically ranked ahead of lower ads, then moving towards this new axiomatic model will increase revenue in expectation.

A drawback of the most general possible formulation is that computing the optimal allocation of ads is unlikely to be computationally feasible because one cannot simply rank the ads by a product of their bid and their quality score.  Instead, one would likely need to try each possible configuration of ads in order to choose the optimal configuration, and this is likely to be too slow to be useful in practice.  For this reason, we also develop a second formulation that is a special case of our most general methodology that has the advantage of admitting a rapidly converging algorithm for computing the optimal allocation of ads.

In addition to presenting methodology to explicitly consider the externalities imposed by one ad on another ad, we also present methodology for capturing the fact that ads from well-known high-quality brands are likely to be less adversely affected by being shown on lower positions on the page than non-brand ads.  In the formulation we present, it is easy to rule out certain configurations of ads as suboptimal, and simply trying each of the non-dominated configurations is likely to enable one to find the optimal allocation of ads sufficiently quickly to be useful in practice.  However, the purely greedy strategy of ranking the ads by their expected cost-per-1000-impressions (eCPM) bids will still not work in this formulation, and may, in fact, lead to lower economic welfare than not considering these brand effects at all.

While a few papers have presented theoretical analyses of circumstances where the click-through rates are not equal to a product of the quality score of an ad and the quality score of a position (\cite{Aggarwal08}, \cite{Athey11}, \cite{Ghosh08}, \cite{Jerath11}, and \cite{Kempe08}), the vast majority of papers on position auctions only consider situations in which click probabilities can be expressed in this manner (\emph{e.g.} \cite{Abrams07}, \cite{Aggarwal06a}, \cite{Ashalgi07}, \cite{Athey12}, \cite{Caragiannis11}, \cite{Cary07}, \cite{Das08}, \cite{Edelman07}, \cite{Edelman10}, \cite{Even-Dar08}, \cite{Feldman07}, \cite{Feldman11}, \cite{Feng07}, \cite{Gomes14}, \cite{Hummel14}, \cite{Kuminov09}, \cite{Lahaie06}, \cite{PaesLeme09}, \cite{PaesLeme10}, \cite{Lucier11}, \cite{Lucier12}, \cite{Roughgarden12}, \cite{Szymanski06}, \cite{Varian07}, \cite{Vorobeychik07}, and \cite{Zhou07}).  Furthermore, \cite{Aggarwal06b} and \cite{Gonen08} have noted that this separability property can be quite important for obtaining desirable properties such as the ability to choose the efficient allocation by greedily ranking of the ads.

In addition, the few papers that explicitly consider non-separable click-through rates differ significantly from our paper.  \cite{Aggarwal08} and \cite{Kempe08} consider models in which users search from the top of the page to the bottom that leads to non-separable click probabilities and tractable algorithms for choosing the optimal allocation of ads, but do not consider more general models of click probabilities, as we do in the present paper.  \cite{Giotis08}, \cite{Gomes09}, and \cite{Roughgarden12} further analyze the equilibrium and efficiency properties of such a model, and \cite{Gomes09} also empirically estimates the parameters in such a model using data from Microsoft Live, but \cite{Jeziorski14} finds empirical evidence that models in which users search from the top of the page to the bottom do not fully match the data. 

Other papers consider different models of externalities, but these still differ from our work in significant ways.  \cite{Athey11} and \cite{Jerath11} both consider alternative models of user behavior that result in non-separable click probabilities, but do not present algorithms for choosing the optimal allocation of ads.  \cite{Fotakis11} and \cite{Ghosh08} address questions related to the computational hardness of the winner-determination problem in separate models of position auctions with externalities with alternative assumptions, but do not present axiomatic methodology for new models of predicted click-through rates or tractable methods for computing the optimal allocation of ads in such models.  Lastly, \cite{Ieong14} presents a model of advertising in a stream in which ads can only be placed in between certain news items that provide a reasonable context for the ad, but does not address the question of externalities that we consider in this paper.

Finally, our paper also relates to general economics literature on auctions with allocative externalities such as \cite{Jehiel96}, \cite{Jehiel01}, \cite{Jehiel96b}, \cite{Jehiel99}, \cite{Katz85}, and \cite{Katz86}.  This literature has considered, amongst other issues, the possibility that a player's payoff might be affected by which of the other bidders wins the auction in addition to whether the player wins the auction.  However, this work has not been applied to Internet position auctions, which is our focus in the present paper.

\section{Model of Externalities}\label{s:model}

There is an auction for $s$ advertising positions on a page.  Each advertising position $k$ has a quality score $n_{k}$, where we assume without loss of generality that $n_{k}$ is non-increasing in $k$.  There are also $m$ advertisers.  Each advertiser $i$ has a quality score $q_{i}$ reflecting the relative clickability of the ad and makes a bid $b_{i}$ reflecting the maximum amount that this advertiser will pay per click.

In this setting, a standard model of position auctions such as \cite{Edelman07} or \cite{Varian07} would assume that the probability advertiser $i$ receives a click in position $k$ is $n_{k} q_{i}$.  We relax this assumption by allowing the probability an advertiser receives a click to depend on these quality scores in a more nuanced way.  To do this, we first identify axioms that we believe should be satisfied by a reasonable model of click probabilities.  We let $p_{(j)} = f_{j}(q_{(1)}, \ldots, q_{(s)}; n_{1}, \ldots, n_{s})$ denote the probability that the advertiser in the $j^{th}$ position receives a click as a function of the quality scores of the ads in the first $s$ positions as well as quality scores of the $s$ positions.  We stipulate that this probability should satisfy these axioms:

(A1)  The probability an ad receives a click is $0$ if either the ad's quality score is $0$ or the ad's position has a quality score of $0$.  That is, $f_{j}(q_{(1)}, \ldots, q_{(s)}; n_{1}, \ldots, n_{s}) = 0$ if either $q_{(j)} = 0$ or $n_{j} = 0$.

(A2)  The probability an ad receives a click is increasing in the ad's quality score.  That is, we require that $f_{j}(q_{(1)}, \ldots, q_{(s)}; n_{1}, \ldots, n_{s})$ is increasing in $q_{(j)}$ for all $j$ satisfying $n_{j} > 0$.

(A3)  The probability an ad in a particular position receives a click is increasing in the quality score of that position.  That is, we require that $f_{j}(q_{(1)}, \ldots, q_{(s)}; n_{1}, \ldots, n_{s})$ is increasing in $n_{j}$ for all $j$ satisfying $q_{(j)} > 0$.

(A4)  The probability an ad in a particular position receives a click is non-increasing in the quality scores of ads in other positions.  That is, $f_{j}(q_{(1)}, \ldots, q_{(s)}; n_{1}, \ldots, n_{s})$ is non-increasing in $q_{(k)}$ for all $k \neq j$. 

(A5)  Increasing the quality score of an ad in a higher quality position decreases the click-through rates of ads in other positions by more than increasing the quality score of an ad in a lower quality position.  Formally, let $\vec{q} \equiv (q_{(1)}, \ldots, q_{(s)})$ denote a vector of qualities for which $q_{(i)} = q_{(k)} = q^{*}$ for some particular $i$ and $k$ satisfying $n_{i} > n_{k}$.  Also let $\vec{q}_{(i)}$ denote the vector of qualities that would result from replacing $q_{(i)} = q^{*}$ with $q_{(i)} = \hat{q}$ for some $\hat{q} \neq q^{*}$, and let $\vec{q}_{(k)}$ denote the vector of qualities that would result from replacing $q_{(k)} = q^{*}$ with $q_{(k)} = \hat{q}$ for the same $\hat{q}$.  Then $|f_{j}(\vec{q}_{(i)}; n_{1}, \ldots, n_{s}) - f_{j}(\vec{q}; n_{1}, \ldots, n_{s})| \geq |f_{j}(\vec{q}_{(k)}; n_{1}, \ldots, n_{s}) - f_{j}(\vec{q}; n_{1}, \ldots, n_{s})|$ for all $j \notin \{i, k\}$.

While properties (A1)-(A3) are intuitive properties that are satisfied by standard models of click-through rates such as \cite{Edelman07} and \cite{Varian07}, axioms (A4) and (A5) are less standard.  Both of these axioms capture the possibility that an ad may impose a negative externality that affects whether other ads receive a click.

In particular, axiom (A4) simply reflects the possibility that when a higher quality ad assumes a particular position, the ad is likely to decrease the probability that ads in other positions receive a click.  This axiom is plausible because if the quality of an ad in a particular position increases, users are relatively more likely to click on this ad, which in turn draws their attention from other ads and decreases the likelihood that users will click on other ads.

Similarly, axiom (A5) reflects the fact that increasing the quality of an ad in a higher quality position does more to decrease the click-through rates of ads in other positions than increasing the quality of an ad in a lower quality position.  Increasing the quality of an ad in a higher quality position does more to increase the probability that users will click on that ad, so increasing the quality of an ad in a higher quality position also draws more user attention from other ads than increasing the quality of an ad in a lower quality position.  Thus both axioms (A4) and (A5) reflect sensible properties on how changing the qualities of ads in other positions is likely to affect the probabilities that other ads receive a click.

Throughout our analysis of models of position auctions, we focus on mechanisms in which the auctioneer chooses the allocation of ads that would maximize total expected welfare with respect to the bids of the advertisers.  That is, the auctioneer chooses the allocation of ads that results in the highest value of $\sum_{j=1}^{s} b_{(j)} p_{(j)}$, where $b_{(j)}$ denotes the cost per click bid of the advertiser in the $j^{th}$ position and $p_{(j)}$ denotes the probability that the advertiser in the $j^{th}$ position receives a click.

In setting prices we focus on a generalization of the generalized second price auction in which the advertiser in the $j^{th}$ position is charged a cost per click $c_{(j)}$ that represents the smallest bid that this advertiser could make while still maintaining the $j^{th}$ position when the allocation of ads is chosen using the above algorithm.  In the special case in which $f_{j}(q_{(1)}, \ldots, q_{(s)}; n_{1}, \ldots, n_{s}) = n_{j} q_{(j)}$ for all $j$, as in \cite{Edelman07} and \cite{Varian07}, this pricing algorithm would indeed result in the same prices selected by the standard generalized second price auction.

\section{General Results}

We first derive some general results on how using an alternative model of predicted click-through rates meeting the axioms given in the previous section would affect revenue from online auctions.  To do this, we compare two otherwise identical methods for predicting the click-through rates of ads in position auctions.  The first method is one in which the predicted click-through rates of the ads in slots $j \notin \{k, k+1\}$ are independent of the quality scores of the ads in positions $k$ and $k+1$, as in a standard model.  The other method we consider is one in which the predicted click-through rates of the ads in positions $j \notin \{k, k+1\}$ may depend on the quality scores of the ads in positions $k$ and $k+1$ in a manner that satisfies the axioms (A4) and (A5) presented in the previous section.  

There are two different ways that incorporating the possibility that the quality scores of ads may affect the click-through rates of ads in other positions could affect revenue.  First there is the possibility that this could affect the allocation of ads that is shown in the auction.  In this case, if the revised model of predicted click-through rates is more accurate, then one would choose a more efficient allocation of ads, and thereby typically achieve higher revenue.

However, in a substantial percentage of auctions, allowing for the possibility that an ad's predicted click-through rate may depend on the quality scores of the other ads will not change the allocation of ads but will affect the pricing.  It is thus important to assess how the prices that the advertisers pay would be affected by the changed model of predicted click-through rates even if this does not affect the allocation of ads.  This is addressed in the following theorem:

\begin{theorem}\label{t:revenue}  Consider two different models of predicted click-through rates for position auctions that are identical except for the following:

(1)  For the first model, the predicted click-through rates of ads in slots $j \notin \{k, k+1\}$ are independent of the quality scores of the ads in slots $k$ and $k+1$.

(2)  For the second model, the predicted click-through rates of ads in slots $j \notin \{k, k+1\}$ depend on the quality scores of the ads in slots $k$ and $k+1$ in a manner that satisfies axioms (A4) and (A5).

Then if the allocation of ads that is selected by the two models of predicted click-through rates is identical, the advertiser in position $k$ pays more per click under the second model if and only if $q_{(k)} > q_{(k+1)}$.

\end{theorem}

All omitted proofs are in the appendix.  Theorem \ref{t:revenue} indicates that if we take into account the externalities that the ads in positions $k$ and $k+1$ impose on the other ads, then the advertiser in position $k$ will pay more per click if and only if this advertiser has a higher quality ad.  Since Theorem \ref{t:revenue} applies to all slots $k$, repeatedly applying Theorem \ref{t:revenue} to every slot suggests that if a model with externalities has no effect on the allocation of ads, then this model will typically increase the cost per click paid by an advertiser if and only if this advertiser's quality score exceeds that of the advertiser just below him.

The results of this section suggest that if one can more accurately describe click probabilities by using a model of the form in Section \ref{s:model}, then one should be able to increase revenue.  Typically higher quality ads will be ranked higher than lower quality ads, so the result in Theorem \ref{t:revenue} suggests that even if this model does not change the allocation of ads, revenue should still increase.  And if one is able to choose a more efficient allocation, then one would also expect revenue to increase.  Thus revenue is likely to increase from using predicted click-through rates of the form in Section \ref{s:model} as long as such a model is more accurate.

\section{Practical Formulation}

For general models of the form in Section \ref{s:model}, it may be difficult to select the efficiency-maximizing configuration because there are an exponentially large number of feasible configurations and it is not obvious how one can rule out different configurations as dominated by others.  Thus it is important to use a model where one can select the efficiency-maximizing configuration in a computationally tractable way.  In this section we present a specific formulation of the model in Section \ref{s:model} that permits such a practical implementation.

In particular, in this section we consider a model in which the predicted click-through rates for the ads in position $i$ are of the form

\[
p_{i} = \frac{\nu n_i q_i}{1 + \lambda \sum_{j=1}^{s} n_j q_j}
\]

\noindent where $\lambda$ and $\nu$ are positive constants, $n_i$ denotes the quality score of the $i^{th}$ position, and $q_i$ denotes the quality score of the ad in the $i^{th}$ position.  This formulation is sensible because one would expect the percentage decrease in an ad's click-through rate due to the negative externalities imposed by the other ads to be proportional to the total click-through rates of these ads, meaning an ad's click-through rate is likely to be decreased by a factor proportional to $1 + \lambda \sum_{j=1}^{s} n_j q_j$.  Also note that in this formulation, setting $\lambda = 0$ and $\nu = 1$ would recover the standard formulation of predicted click-through rates, so optimally choosing these parameters can never result in less accurate predicted click-through rates than the standard formulation.  Further note that changing the value of $\nu$ would never change the optimal allocation of ads.  Here $\nu$ is a term that only serves to make the predicted click-through rates unbiased on average.

Now define $S$ to be the expected social welfare from a given ranking of ads when $\nu = 1$.  That is define $S$ to be what is given by the following formulation:

\begin{equation}\label{eq:welfare}
S = \frac{\sum_{i=1}^{s} n_i b_i q_i}{1 + \lambda \sum_{j=1}^{s} n_j q_j}
\end{equation}

\noindent where $b_i$ denotes the bid of the advertiser in the $i^{th}$ position.  We first begin our analysis of this formulation by noting when using a non-zero value of $\lambda$ would result in changing the allocation of ads:

\begin{theorem}\label{t:switch}  Welfare is enhanced by switching the order of the ads in positions $k$ and $m$ where $k < m$ if and only if $b_{m} q_{m} - \lambda q_{m} S > b_{k} q_{k} - \lambda q_{k} S$.

\end{theorem}

\proof  Let $S_{k,m}$ denote the expected social welfare that would arise from switching the order of the ads in positions $k$ and $m$.  Note that

\[
S_{k,m} = \frac{\sum_{i=1}^{s} n_i b_i q_i + (n_k - n_m)(b_m q_m - b_k q_k)}{1 + \lambda \sum_{j=1}^{s} n_j q_j + \lambda (n_k - n_m)(q_m - q_k)}
\]

From this it follows that $S_{k,m} > S$, where $S$ is the social welfare that would arise from the original ordering of the ads, if and only if

\begin{eqnarray*}
\bigg(\sum_{i=1}^{s} && n_i b_i q_i + (n_k - n_m)(b_m q_m - b_k q_k)\bigg)\bigg(1 + \lambda \sum_{j=1}^{s} n_j q_j\bigg) > \\ && \bigg(\sum_{i=1}^{s} n_i b_i q_i\bigg)\bigg(1 + \lambda \sum_{j=1}^{s} n_j q_j + \lambda (n_k - n_m)(q_m - q_k)\bigg)
\end{eqnarray*}

\noindent which in turn holds if and only if

\[
(n_k - n_m)(b_m q_m - b_k q_k)\bigg(1 + \lambda \sum_{j=1}^{s} n_j q_j\bigg) > \bigg(\sum_{i=1}^{s} n_i b_i q_i\bigg)\lambda (n_k - n_m)(q_m - q_k).
\]

\noindent  Now since $k < m$, we have $n_k - n_m > 0$, and the above inequality holds if and only if

\[
(b_m q_m - b_k q_k)\bigg(1 + \lambda \sum_{j=1}^{s} n_j q_j\bigg) > \bigg(\sum_{i=1}^{s} n_i b_i q_i\bigg)\lambda (q_m - q_k)
\]

\noindent which in turn holds if and only if $b_m q_m - b_k q_k > S \lambda (q_m - q_k)$.  Thus welfare is enhanced by switching the order of the ads in positions $k$ and $m$ where $k < m$ if and only if $b_{m} q_{m} - \lambda q_{m} S > b_{k} q_{k} - \lambda q_{k} S$.  \qed  \endproof

Theorem \ref{t:switch} suggests that if one can obtain a good estimate of the social welfare $S$ that will result in the efficiency-maximizing configuration, then it may be feasible to rank the ads on the basis of scores of the form $b_{m} q_{m} - \lambda q_{m} S$ to achieve the efficiency maximizing allocation.  We now exploit this insight to derive a computationally efficient way of selecting the optimal ordering of ads.

The algorithm proceeds by selecting a value $S_{L}$ that is lower than the social welfare $S$ that will result in the efficiency-maximizing configuration and another value $S_{H}$ that is higher than the social welfare $S$ that will result in the efficiency-maximizing configuration.  The algorithm then repeatedly replaces either $S_{L}$ or $S_{H}$ with $\hat{S} \equiv \frac{1}{2}(S_L + S_H)$ until it finds some such value of $\hat{S}$ that is guaranteed to result in the efficiency-maximizing allocation when ranking ads on the basis of scores of the form $b_{m} q_{m} - \lambda q_{m} S$ when $S = \hat{S}$.  Such an algorithm will typically require very few steps in practice because after $n$ passes, we know that $\hat{S}$ will be within a factor of $2^{-n}$ of the true social welfare $S$ corresponding to the efficiency-maximizing configuration.  The detailed steps for the algorithm are as follows:

(1)  Define $S_{L}$ to be the expected social welfare calculated in equation (\ref{eq:welfare}) that would result if the ads were ranked on the basis of the scores $b_{m} q_{m}$.

(2)  Define $S_{H} \equiv \sum_{m=1}^{s} n_{m} b_{m} q_{m}$ when the ads are ranked on the basis of the scores $b_{m} q_{m}$.

(3)  Calculate the rankings of the ads when the ads are ranked on the basis of the scores $b_{m} q_{m} - \lambda q_{m} S$ for $S = S_{L}$ and $S = S_{H}$.

(4)  If the rankings of the ads in step (3) are the same for both $S = S_{L}$ and $S = S_{H}$, then choose this ranking of the ads.

(5)  If these rankings are different, let $\hat{S} \equiv \frac{1}{2}(S_L + S_H)$ and calculate the ranking of the ads when the ads are ranked on the basis of the scores $b_{m} q_{m} - \lambda q_{m} \hat{S}$.

(6)  Let $\phi(\hat{S}) \equiv \sum_{m=1}^{s} n_{m} (b_{m} q_{m} - \lambda q_{m} \hat{S})$ when the ads are ranked on the basis of the scores $b_{m} q_{m} - \lambda q_{m} \hat{S}$.  If $\phi(\hat{S}) < \hat{S}$, then let $S_{H} = \hat{S}$.  Otherwise let $S_{L} = \hat{S}$.

(7)  Repeat steps (3)-(6) until the rankings in step (4) are the same for both $S = S_{L}$ and $S = S_{H}$, and choose the resulting ranking of ads.

This algorithm indeed results in the efficiency-maximizing allocation:

\begin{theorem}\label{t:algorithm}  The ranking of ads that results from the algorithm considered above is the efficiency-maximization allocation.

\end{theorem}

\proof Define $\phi(S)$ to be equal to the value of $\sum_{m=1}^{s} n_{m} (b_{m} q_{m} - \lambda q_{m} S)$ that would result when the ads are ranked on the basis of the scores $b_{m} q_{m} - \lambda q_{m} S$.  Note that if $S$ is the social welfare calculated in equation (\ref{eq:welfare}) for the efficiency-maximizing allocation, then $\phi(S) = S$ because $\phi(S) = \sum_{m=1}^{s} n_{m} (b_{m} q_{m} - \lambda q_{m} S) = (\sum_{m=1}^{s} n_{m} b_{m} q_{m}) (1 - \lambda \frac{\sum_{m=1}^{s} n_{m} q_{m}}{1 + \lambda \sum_{j=1}^{s} n_{j} q_{j}}) = (\sum_{m=1}^{s} n_{m} b_{m} q_{m}) (\frac{1}{1 + \lambda \sum_{j=1}^{s} n_{j} q_{j}}) = S$.  Also note that $\phi(S)$ is decreasing in $S$, so if $S^{*}$ is the social welfare calculated in equation (\ref{eq:welfare}) for the efficiency-maximizing allocation, then $\phi(S) < S$ when $S > S^{*}$ and $\phi(S) > S$ when $S < S^{*}$.

Now note that the value of $S_{L}$ computed in step (1) of the above algorithm is necessarily no greater than the social welfare that would result in the efficiency-maximizing allocation because this is the social welfare resulting from some feasible allocation of ads.  Also note that the value of $S_{H}$ computed in step (2) of the above algorithm is necessarily greater than the social welfare that would result in the efficiency-maximizing allocation because the value of $\frac{\sum_{i=1}^{s} n_i b_i q_i}{1 + \lambda \sum_{j=1}^{s} n_j q_j}$ is bounded above by $\sum_{i=1}^{s} n_i b_i q_i$, which is in turn bounded above by the value of $S_{H}$ in step (2).  From this it follows that if $S^{*}$ is the social welfare calculated in equation (\ref{eq:welfare}) for the efficiency-maximizing allocation, then $S_{L} \leq S^{*} < S_{H}$.

Now if the rankings of the ads that are calculated in step (3) of the above algorithm are the same for both $S = S_{L}$ and $S = S_{H}$, then the ranking of the ads that would result from ranking the ads on the basis of scores of the form $b_{m} q_{m} - \lambda q_{m} S$ when $S = S^{*}$ is the same as the rankings of the ads that result when ranking the ads on the basis of such scores with either $S = S_{L}$ or $S = S_{H}$.  But the ranking that results from ranking the ads on the basis of scores of the form $b_{m} q_{m} - \lambda q_{m} S^{*}$ is the efficiency maximizing allocation, so from this it follows that if the rankings of the ads that are calculated in step (3) of the above algorithm are the same for both $S = S_{L}$ and $S = S_{H}$, then choosing this ranking indeed results in the efficiency-maximizing allocation.

If these rankings are different, then we know that $\phi(S_L) \geq S_L$ and $\phi(S_H) < S_H$ since $S_{L} \leq S^{*} < S_{H}$ and if $S^{*}$ is the social welfare calculated in equation (\ref{eq:welfare}) for the efficiency-maximizing allocation, then $\phi(S) < S$ when $S > S^{*}$ and $\phi(S) > S$ when $S < S^{*}$.  Furthermore, after steps (5) and (6) are executed, it will be the case that $\phi(S_L) \geq S_L$ and $\phi(S_H) < S_H$.  From this it follows that after steps (5) and (6) are executed, it will still be the case that $S_{L} \leq S^{*} < S_{H}$ when $S^{*}$ is the social welfare in equation (\ref{eq:welfare}) for the efficiency-maximizing allocation.

But this means that if the rankings of the ads calculated in step (3) of the above algorithm after executing steps (5) and (6) are the same for both $S = S_{L}$ and $S = S_{H}$, then this ranking would result in the efficiency-maximizing allocation by the same reasoning as in the third paragraph of this proof.  Furthermore, if these rankings are not the same, then repeating the process in steps (3)-(6) will ultimately result in values of $S_{L}$ and $S_{H}$ for which these rankings are the same because the values of $S_{L}$ and $S_{H}$ will eventually get arbitrarily close to one another.  From this it follows that the ranking of ads that results from the algorithm considered above is the efficiency-maximization allocation.  \qed  \endproof

Having presented results for a model of position auctions with externalities, we now turn attention to an alternative model in which there are no externalities but an ad's click probability still cannot be expressed as the product of a quality score of a position and a quality score of an advertiser.  In particular, we consider a model in which the click-through rates of some advertisers are more sensitive to the precise position in which they are displayed than the click-through rates of other advertisers.  This possibility has been demonstrated to hold empirically, as \cite{Jerath11} has noted that the click-through rates of brand ads are less adversely affected by being displayed in low positions than those of non-brand ads.

\section{Model of Brand Effects}\label{s:model2}

There is an auction for $s$ advertising positions involving $m$ advertisers.  Some advertisers are brand advertisers and other advertisers are non-brand advertisers.  Each advertiser $i$ has a quality score $q_{i}$ reflecting the relative click-through rate of the ad if the ad is in the top position.  Each advertiser $i$ also makes a bid $b_{i}$ reflecting the maximum amount that this advertiser will pay per click.

How an advertiser's probability of receiving a click varies with the advertiser's position on the page depends on whether the advertiser is a brand advertiser or a non-brand advertiser.  Each advertising position $k$ has two separate quality scores $\beta_{k}$ and $\eta_{k}$, where we assume that both $\beta_{k}$ and $\eta_{k}$ are non-increasing in $k$.  If a brand (non-brand) advertiser with quality score $q$ is shown in position $k$, then the advertiser's click probability is $\beta_{k} q$ ($\eta_{k} q$).  We normalize the values of $\beta_{k}$ and $\eta_{k}$ by assuming without loss of generality that $\beta_{1} = \eta_{1} = 1$.

Our model of position auctions with brand effects allows for the possibility that some advertisers may be less adversely affected by having their ads shown in lower positions than other advertisers.  In particular, if $\beta_{k}$ does not decline as rapidly with $k$ as $\eta_{k}$, then brand advertisers would be less adversely affected by being displayed in a lower position than non-brand advertisers, as is the case empirically \cite{Jerath11}.  This model also encompasses the standard model of position auctions considered in \cite{Edelman07} and \cite{Varian07} as a special case in which $\beta_{k} = \eta_{k}$ for all $k$.  

As in our analysis of models of position auctions with externalities, we focus on mechanisms in which the auctioneer seeks to maximize total expected welfare with respect to the bids of the advertisers.  That is, the auctioneer chooses the allocation of ads that maximizes $\sum_{j=1}^{s} b_{(j)} p_{(j)}$, where $b_{(j)}$ denotes the bid of the advertiser in the $j^{th}$ position and $p_{(j)}$ denotes the probability that the ad in the $j^{th}$ position is clicked.

\section{Optimal Allocation}

We begin our analysis by noting some simple properties about the optimal allocation of ads with brand effects.  While it is difficult to give a full characterization of the ordering of all the ads, we can present a partial ordering of the positions the ads will assume by noting how one should order the brand ads amongst each other as well as the non-brand ads amongst each other:

\begin{observation}\label{o:brand-monotone}  Let $B$ denote the set of positions in which brand ads are displayed and let $N$ denote the set of positions in which non-brand ads are displayed.  Then the optimal allocation of ads involves showing the brand ad with the highest eCPM in the highest position in $B$, showing the brand ad with the second-highest eCPM in the second-highest position in $B$, and in general showing the brand ad with the $k^{th}$-highest eCPM in the $k^{th}$-highest position in $B$.  Similarly, the optimal allocation of ads involves showing the non-brand ad with the highest eCPM in the highest position in $N$, showing the non-brand ad with the second-highest eCPM in the second-highest position in $N$, and in general showing the non-brand ad with the $k^{th}$-highest eCPM in the $k^{th}$-highest position in $N$.

\end{observation}

A consequence of this result is that, given a decision about which positions to show brand ads and non-brand ads in, the optimal allocation for this selection is uniquely determined by ordering the brand ads by their eCPM bids and doing the same for the non-brand ads.  Thus in order to determine the optimal allocation of ads in the model in Section \ref{s:model2}, it is only necessary to check each possible configuration of positions in which one can show brand ads, as there is only one possible allocation that may be optimal given this configuration.  Since there are typically only a small number of slots in most position auctions, checking each of these possibilities is something that is quite computationally feasible.

Given the above result, the only interesting question is which positions the auctioneer should show a brand ad in.  While a full characterization of this precise allocation is intractable, we can present such a characterization for some special cases.  We first show that if the brand ads are not adversely affected by being shown in the lower positions, then the optimal allocation will involve showing non-brand ads in the top positions and brand ads in the lower positions:

\begin{theorem}\label{t:brand-last}  If $\beta_{k}$ is independent of $k$ for all $k$, but $\eta_{k}$ is strictly decreasing in $k$, then the optimal allocation involves not showing any non-brand ads in a lower position than a brand ad.

\end{theorem}

While Theorem \ref{t:brand-last} presents a precise description of the nature of the optimal allocation ads in an important special case of this model, it is also important to ask how one would choose the allocation of ads in more general cases.  Although we do not present a full characterization of the optimal allocation of ads for general values of the separate quality scores for the positions, we can present the following partial characterization of the circumstances under which the auctioneer will want to show a brand ad in the next position.  

\begin{theorem}\label{t:partial}  For a fixed allocation of ads in the first $k$ positions, a fixed set of eCPM bids for the remaining non-brand advertisers, and a fixed set of eCPM bids for all the remaining brand advertisers except for the brand advertiser with the highest eCPM bid amongst the remaining brand advertisers, either it is never optimal to show a brand ad in the $k+1^{th}$ position, always optimal, or it is optimal to show a brand ad in the $k+1^{th}$ position if and only if this advertiser's eCPM bid exceeds some threshold.

\end{theorem}

\proof Let $v_{b}$ denote the eCPM bid of the highest remaining brand advertiser, let $v_{n}$ denote the eCPM bid of the highest remaining non-brand advertiser, let $V_{B}$ denote the maximum total expected welfare that could be obtained from the advertisers in positions $k+2$ to $s$ if the auctioneer displays a brand ad in position $k+1$, and let $V_{N}(v_{b})$ denote the maximum total expected welfare that could be obtained from the advertisers in positions $k+2$ to $s$ if the auctioneer displays a non-brand ad in position $k+1$ for any given $v_{b}$.  Note that if $v^{\prime} > v$, then $V_{N}(v^{\prime}) - V_{N}(v) \leq \beta_{k+2} (v^{\prime} - v)$ since increasing the eCPM bid of an eCPM bidder who will not be displayed before the $k+2^{nd}$ position cannot increase the total expected welfare arising from positions $k+2$ to $s$ by more than $\beta_{k+2}$ times the size of this increase.

Now if $b_{(j)}$ denotes the cost per click bid of the advertiser in the $j^{th}$ position and $p_{(j)}$ denotes the probability that the advertiser in the $j^{th}$ position receives a click, then total expected welfare if the auctioneer displays a brand ad in position $k+1$ is $\sum_{j=1}^{k} b_{(j)} p_{(j)} + \beta_{k+1} v_{b} + V_{B}$ and total expected welfare if the auctioneer displays a non-brand ad in position $k+1$ is $\sum_{j=1}^{k} b_{(j)} p_{(j)} + \eta_{k+1} v_{n} + V_{N}(v_{b})$.  From this it follows that the auctioneer should display a brand ad in position $k+1$ if and only if $\sum_{j=1}^{k} b_{(j)} p_{(j)} + \beta_{k+1} v_{b} + V_{B} \geq \sum_{j=1}^{k} b_{(j)} p_{(j)} + \eta_{k+1} v_{n} + V_{N}(v_{b})$, which holds if and only if $\beta_{k+1} v_{b} - V_{N}(v_{b}) \geq \eta_{k+1} v_{n} - V_{B}$.

But since $v^{\prime} > v$ implies that $V_{N}(v^{\prime}) - V_{N}(v) \leq \beta_{k+2} (v^{\prime} - v)$ and $\beta_{k+1} \geq \beta_{k+2}$, it follows that $\beta_{k+1} v_{b} - V_{N}(v_{b})$ is non-decreasing in $v_{b}$.  From this it follows that either $\beta_{k+1} v_{b} - V_{N}(v_{b}) \geq \eta_{k+1} v_{n} - V_{B}$ never holds for all $v_{b}$, always holds for all $v_{b}$, or holds if and only if $v_{b}$ is greater than or equal to some threshold.  Thus either it is never optimal to show a brand ad in the $k+1^{th}$ position, always optimal, or it is optimal to show a brand ad in the $k+1^{th}$ position if and only if this advertiser's eCPM bid exceeds some threshold.  \qed  \endproof

It is worth noting that while it will be optimal to show a brand ad in the next-highest position if and only if the eCPM of the brand ad exceeds some threshold, in general this threshold need not be equal to the highest eCPM of the remaining non-brand ads.  We analyze the consequences of this greedy approach to choosing the allocation of ads in the next section.

\section{Greedy Allocation}

In the standard model of position auctions without brand effects, one simply ranks the ads in order of their estimated eCPMs by showing the ad with the highest eCPM in the highest position, the ad with the second-highest eCPM in the second position, and in general showing the ad with the $k^{th}$-highest eCPM in the $k^{th}$ position.  A natural analog of this approach for the model with brand effects would be to show the ad with highest normalized eCPM in the first position, show the ad with the highest normalized eCPM amongst all remaining ads in the second position, and in general show the ad with the highest normalized eCPM amongst all ads not in the first $k-1$ positions in the $k^{th}$ position, where the normalized eCPM for advertiser $i$ in the $k^{th}$ position is equal to $\beta_{k} b_{i} q_{i}$ if the ad is a brand ad and $\eta_{k} b_{i} q_{i}$ if the ad is a non-brand ad.

While the greedy strategy is suitable in a standard position auction framework without brand effects, when there are brand effects, such a strategy may only result in a fraction of the maximum possible efficiency that could be achieved:

\begin{theorem}\label{t:greedy-half}  Suppose the auctioneer uses the greedy strategy to choose the efficient allocation.  Such an algorithm is guaranteed to achieve at least half of the maximum possible total expected welfare.  Moreover, this bound is tight.

\end{theorem}

\proof Note that if the greedy strategy fails to select the efficient allocation, then there must exist some position $j$ for which the greedy strategy elects to display an ad that would not be the best ad to display in that position in the optimal allocation.  Thus in the optimal allocation, it must be best to instead display an ad with a lower normalized eCPM in position $j$ and display the ad that the greedy strategy would display in position $j$ in some other position $k > j$.

But this means that the normalized eCPM of the ad that is displayed by the greedy strategy in position $j$ is necessarily greater than the normalized eCPM of both the ad that the optimal allocation would display in position $j$ as well as the normalized eCPM of the ad that the optimal allocation would display in position $k$.  Thus the sum of the normalized eCPMs of the ads that are displayed in positions $j$ and $k$ by the greedy strategy is necessarily at least half as large as the sum of the normalized eCPMs of the ads that are displayed in positions $j$ and $k$ in the optimal allocation.

Since the above result holds for any position $j$ where the greedy strategy displays a different ad than the ad would be best to display in that position under the optimal allocation, it then follows that the greedy strategy can never result in the loss of more than half of the maximum possible total expected welfare that can be achieved.  This proves the first part of the theorem.

To see that the lower bound is tight, suppose there are $s = 2$ positions, one brand advertiser with eCPM bid $b_{i} q_{i} = 1 + \epsilon$ for some small $\epsilon > 0$, and one non-brand advertiser with eCPM bid $b_{i} q_{i} = 1$.  Also suppose that $\beta_{2} = \beta_{1} = \eta_{1} = 1$ and $\eta_{2} = 0$.  In this case, the purely greedy strategy will display the brand ad in the first position and the non-brand ad in the second position, resulting in a total expected welfare of $1 + \epsilon$.  However, the optimal strategy will display the non-brand ad in the first position and the brand ad in the second position, resulting in a total expected welfare of $2 + \epsilon$.  Since $\lim_{\epsilon \to 0} \frac{1 + \epsilon}{2 + \epsilon} = \frac{1}{2}$, it follows that one cannot improve on the bounds given in the previous paragraph.  Thus the bound given in the statement of the theorem is tight.  \qed  \endproof

Theorem \ref{t:greedy-half} indicates that using the greedy strategy to choose the allocation of ads can result in significant efficiency losses.  The reason for this is that the greedy strategy will sometimes show a brand ad that is only marginally better than a non-brand ad in a higher position even though that brand ad would be significantly better than the non-brand ad in the lower position.  In these situations, following the greedy strategy will result in a significantly less efficient allocation than showing the brand ad in the lower position.

While the greedy strategy may fail to obtain a significant percentage of the possible efficiency gains that could be achieved, one might conjecture that this strategy would still result in higher economic welfare than not taking brand effects into account at all.  However, even using an auction framework without brand effects can still outperform the greedy strategy when there are brand effects, as the following theorem illustrates:

\begin{theorem}\label{t:greedy-standard}  The greedy strategy can result in lower expected welfare than selecting the allocation that would be selected by the standard model of predicted click-through rates without brand effects.

\end{theorem}

\proof  Suppose there are $s = 3$ positions, there is one brand advertiser with an eCPM bid of $b_{i} q_{i} = 10$, a second brand advertiser with an eCPM bid of $b_{i} q_{i} = 1$, and a non-brand advertiser with an eCPM bid of $b_{i} q_{i} = 1 + \epsilon$ for some small $\epsilon > 0$.  Also suppose that $\beta_{3} = \beta_{2} = \beta_{1} = \eta_{1} = 1$, $\eta_{2} = \frac{1}{2}$, and $\eta_{3} = 0$.

Now if the allocation is selected using the standard model of predicted click-through rates without brand effects, then the non-brand advertiser will be ranked ahead of the second brand advertiser, and total expected welfare will be $\beta_{1} 10 + \eta_{2} (1 + \epsilon) + \beta_{3} 1 = 11 \frac{1}{2} + \frac{\epsilon}{2}$.  But if the allocation is selected by the greedy strategy using the model with brand effects, then the second brand advertiser will be ranked ahead of the non-brand advertiser, and total expected welfare will be $\beta_{1} 10 + \beta_{2} 1 + \eta_{3} (1 + \epsilon) = 11 < 11 \frac{1}{2} + \frac{\epsilon}{2}$.  Thus the greedy strategy can result in lower expected welfare than selecting the allocation that would be selected by the standard model of predicted click-through rates without brand effects.  \qed  \endproof

The results in this section indicate that the greedy strategy is a poor choice for choosing the allocation of ads in a model of brand effects since it will frequently lead to significant efficiency losses.  Fortunately, we know from the previous section that there is a simple algorithm for choosing the optimal allocation of ads with brand effects, so following this greedy strategy is unnecessary.

\section{Conclusions}

This paper has presented analysis of alternative models of predicted click-through rates in position auctions that would model both the externalities that ads in some positions impose on ads in other positions as well as the possibility that certain ads would be less adversely affected by being displayed in lower positions than others.  We have presented general axioms that such models might satisfy, analyzed the revenue consequences of using these schemes, and considered various algorithms for implementing the selection of ads when the predicted click-through rates are governed by the relationships we have described.

A natural extension of our analysis would be to analyze how bidders' equilibrium strategies would be affected if the true click-through rates were governed by the alternative models of predicted click-through rates in this paper and one moved from using a standard model of predicted click-through rates to using the correct model.  Such analysis would be quite complicated because it would require one to understand both the nature of the inaccurate quality scores that are used when fitting the standard model of predicted click-through rates and then understanding how advertisers would react to a system that uses this inaccurate model of predicted click-through rates given the true model.  Further analysis could reveal how these changes would affect equilibrium strategies.

\newpage

\section*{Appendix - Omitted Proofs}

\textbf{Proof of Theorem \ref{t:revenue}:}  Note that in the second model the predicted click-through rates of the ads in slots $j \notin \{k, k+1\}$ will be higher if the allocation of the ads in positions $k$ and $k+1$ is reversed if and only if $q_{(k)} > q_{(k+1)}$.  To see this, let $\underline{q} \equiv \min \{q_{(k)}, q_{(k+1)}\}$ and let $\overline{q} \equiv \max \{q_{(k)}, q_{(k+1)}\}$.  We know from axiom (A4) that if $f_{j}(q_{(k)}, q_{(k+1)})$ denotes the click-through rate of the ad in position $j \notin \{k, k+1\}$ for given quality scores of the ads in positions $k$ and $k+1$ that $f_{j}(\overline{q}, \underline{q}) < f_{j}(\underline{q}, \underline{q})$ and $f_{j}(\underline{q}, \overline{q}) < f_{j}(\underline{q}, \underline{q})$.  Furthermore, we know from axiom (A5) that $|f_{j}(\overline{q}, \underline{q}) - f_{j}(\underline{q}, \underline{q})| > |f_{j}(\underline{q}, \overline{q}) - f_{j}(\underline{q}, \underline{q})|$.  By combining these facts we see that $f_{j}(\overline{q}, \underline{q}) < f_{j}(\underline{q}, \overline{q})$ for all $j \notin \{k, k+1\}$.  Thus in the second model the predicted click-through rates of the ads in slots $j \notin \{k, k+1\}$ will be higher if the allocation of the ads in positions $k$ and $k+1$ is reversed if and only if $q_{(k)} > q_{(k+1)}$.

Now let $S(q_{(k)}, q_{(k+1)})$ denote the total social welfare that the ads not in positions $k$ and $k+1$ would receive for given quality scores of the ads in positions $k$ and $k+1$ (\emph{i.e.} let $S(q_{(k)}, q_{(k+1)}) \equiv \sum_{j \notin \{k, k+1\}} b_{(j)} p_{(j)} (q_{(k)}, q_{(k+1)})$ where $p_{(j)} (q_{(k)}, q_{(k+1)})$ denotes the predicted click-through rate of an ad in position $j$ for given quality scores of the ads in positions $k$ and $k+1$).  Then total social welfare that arises for given quality scores of the ads in positions $k$ and $k+1$ is $b_{(k)} p_{(k)}(q_{(k)}, q_{(k+1)}) + b_{(k+1)} p_{(k+1)}(q_{(k)}, q_{(k+1)}) + S(q_{(k)}, q_{(k+1)})$.  Thus if the order of the ads in positions $k$ and $k+1$ is reversed, then total social welfare will be $b_{(k+1)} p_{(k)}(q_{(k+1)}, q_{(k)}) + b_{(k)} p_{(k+1)}(q_{(k+1)}, q_{(k)}) + S(q_{(k+1)}, q_{(k)})$.  From this it follows that the advertiser in position $k$ pays a cost per click $c$ satisfying $c p_{(k)}(q_{(k)}, q_{(k+1)}) + b_{(k+1)} p_{(k+1)}(q_{(k)}, q_{(k+1)}) + S(q_{(k)}, q_{(k+1)}) = b_{(k+1)} p_{(k)}(q_{(k+1)}, q_{(k)}) + b_{(k)} p_{(k+1)}(q_{(k+1)}, q_{(k)}) + S(q_{(k+1)}, q_{(k)})$ or $c p_{(k)}(q_{(k)}, q_{(k+1)}) = b_{(k+1)} p_{(k)}(q_{(k+1)}, q_{(k)}) + b_{(k)} p_{(k+1)}(q_{(k+1)}, q_{(k)}) - b_{(k+1)} p_{(k+1)}(q_{(k)}, q_{(k+1)}) + S(q_{(k+1)}, q_{(k)}) - S(q_{(k)}, q_{(k+1)})$.

From this it follows that the advertiser in position $k$ pays a cost per click $c$ that is increasing in $S(q_{(k+1)}, q_{(k)}) - S(q_{(k)}, q_{(k+1)})$.  Now in the first model we know that $S(q_{(k+1)}, q_{(k)}) - S(q_{(k)}, q_{(k+1)}) = 0$ because the predicted click-through rates of the ads in slots $j \notin \{k, k+1\}$ are independent of the quality scores of the ads in positions $k$ and $k+1$, so total social welfare of the ads not in positions $k$ and $k+1$ is also independent of the quality scores of the ads in positions $k$ and $k+1$.  And in the second model, when $q_{(k)} > q_{(k+1)}$, we have $S(q_{(k)}, q_{(k+1)}) < S(q_{(k+1)}, q_{(k)})$ because we know from the first paragraph of this proof that the click-through rates of the ads in slots $j \notin \{k, k+1\}$ will be higher if the allocation of ads in positions $k$ and $k+1$ is reversed when $q_{(k)} > q_{(k+1)}$ so the total social welfare of the ads not in positions $k$ and $k+1$ will also be higher if the allocation of ads in positions $k$ and $k+1$ is reversed.  

By combining these results it follows that if $q_{(k)} > q_{(k+1)}$, then the advertiser in position $k$ pays a higher cost per click in the second model than under the first model.  Similar reasoning then shows that if $q_{(k)} < q_{(k+1)}$, then the advertiser in position $k$ pays a lower cost per click in the second model than under the first model.  From this it follows that if the allocation of ads that is selected by the two models of predicted click-through rates is identical, the advertiser in position $k$ pays more per click under the second model if and only if $q_{(k)} > q_{(k+1)}$.  \qed

\bigskip

\noindent \textbf{Proof of Observation \ref{o:brand-monotone}:}  If $B$ denotes the set of positions in which brand ads are displayed and $N$ denotes the set of positions in which non-brand ads are displayed, then total expected welfare with respect to the bids of the advertisers is $\sum_{j \in B} b_{(j)} p_{(j)} + \sum_{j \in N} b_{(j)} p_{(j)}$, where $b_{(j)}$ denotes the cost per click bid of the advertiser in the $j^{th}$ position and $p_{(j)}$ denotes the probability that the advertiser in the $j^{th}$ position receives a click.  Since $p_{(j)} = q_{(j)} \beta_{j}$ for brand advertisers and $p_{(j)} = q_{(j)} \eta_{j}$ for non-brand advertisers, where $q_{(j)}$ denotes the quality score of the advertiser in the $j^{th}$ position, it then follows that total expected welfare can also be written as $\sum_{j \in B} b_{(j)} q_{(j)} \beta_{j} + \sum_{j \in N} b_{(j)} q_{(j)} \eta_{j}$.

But since $\beta_{j}$ is non-increasing in $j$, it follows that $\sum_{j \in B} b_{(j)} q_{(j)} \beta_{j}$ is maximized by placing the brand ad with the highest value of $b_{i} q_{i}$ in the highest position in $B$, placing the brand ad with the second-highest value of $b_{i} q_{i}$ in the second-highest position in $B$, and in general placing the brand ad with the $k^{th}$-highest value of $b_{i} q_{i}$ in the $k^{th}$-highest position in $B$.  Similarly, $\sum_{j \in N} b_{(j)} q_{(j)} \eta_{j}$ is maximized by placing the non-brand ad with the highest value of $b_{i} q_{i}$ in the highest position in $N$, placing the non-brand ad with the second-highest value of $b_{i} q_{i}$ in the second-highest position in $N$, and in general placing the non-brand ad with the $k^{th}$-highest value of $b_{i} q_{i}$ in the $k^{th}$-highest position in $N$.  From this it follows that the optimal allocation of ads indeed involves following the allocation given in the statement of the observation.  \qed

\bigskip

\noindent \textbf{Proof of Theorem \ref{t:brand-last}:} Suppose by means of contradiction that the optimal allocation involves showing a non-brand ad in a lower position than a brand ad, and in particular, that this allocation contains a brand ad with eCPM bid $v_{b}$ in position $j$, while a non-brand ad with eCPM bid $v_{n}$ is in position $k$, where $j < k$.  Then the total social welfare arising from the ads in position $j$ and $k$ is $\beta_{j} v_{b} + \eta_{k} v_{n}$.  However, the social welfare arising from the ads in these positions would be $\beta_{k} v_{b} + \eta_{j} v_{n} = \beta_{j} v_{b} + \eta_{j} v_{n} > \beta_{j} v_{b} + \eta_{k} v_{n}$ if the order of these ads were switched, meaning it would be possible to choose an alternative allocation of ads with higher social welfare.  This contradicts our original assumption and proves the desired result.  \qed  \endproof

\end{document}